\documentclass[12pt,a4paper]{article}

\usepackage{amssymb}
\usepackage{amsmath}
\usepackage{enumerate}
\usepackage{bm}
\usepackage{latexsym}
\usepackage{epsfig}
\usepackage[utf8]{inputenc}
\usepackage[russian,english]{babel}
\usepackage{array}
\usepackage{footmisc}

\def\J{\mathcal{J}}

\def\dRM{{\mathrm d}}

\def\mx{{\bm x}}
\def\mv{{\bm v}}

\def\mk{{\bm k}}

\def\eps{\varepsilon}
\allowdisplaybreaks
\allowbreak
\def\boldnabla{{\bm \nabla}}
\newcommand{\fp}[1]{FP$_{\textrm{#1}}$}

\begin{document}
\textwidth=135mm
 \textheight=200mm
\begin{center}
{\bfseries Active-to-absorbing phase transition subjected to the velocity fluctuations in the frozen limit case}
\vskip 5mm
N.~V.~Antonov$^{1}$, M. Hnatich$^{2,3,4}$, A.~S.~Kapustin$^{1}$, 
 T. Lu\v{c}ivjansk\'y$^{3,5}$\footnote[2]{corresponding author, email:tomas.lucivjansky@upjs.sk} and L.~Mi\v{z}i\v{s}in$^{3,4}$
\vskip 5mm
{\small {\it $^1$  Department of Theoretical Physics, St. Petersburg University,Ulyanovskaya 1, 
St. Petersburg, Petrodvorets, 198504 Russia}} \\
{\small {\it $^2$ Institute of Experimental Physics, Slovak Academy of Sciences, Watsonova 47, 040 01 Ko\v{s}ice, Slovakia}} \\ 
{\small {\it $^3$  Faculty of Sciences, P.J. \v{S}af\'arik University, Moyzesova 16, 040 01 Ko\v{s}ice, Slovakia}} \\
{\small {\it $^4$  Bogoliubov Laboratory of Theoretical Physics, JINR, 141980 Dubna, Moscow Region, Russia}} \\
{\small {\it $^5$ Fakult\"at f\"ur Physik, Universit\"at Duisburg-Essen, Lotharstra{\ss}e 1, D-47048 Duisburg, Germany}}
\\
\end{center}
\vskip 5mm
\centerline{\bf Abstract}
The directed bond percolation process is studied in the presence of
 compressible velocity fluctuations with long-range correlations.
 We discuss a construction of a field theoretic action and a way of
 obtaining its large scale properties using the perturbative renormalization group.
 The most interesting results for the frozen velocity limit are given.
\vskip 10mm
\selectlanguage{russian}
\centerline{\bf Aннотация}
Исследуется процесс прямой перколяции в присутствии сжимаемых флуктуаций скорости
с крупономасштабными  корреляциями. Обсуждается построение теоретико-полевого действия 
и способ получения его  свойств в области больших масштабов с использованием пертурбативной 
ренормализационной группы. Представлены наиболее интересные результаты в пределе замороженной турублентности.
\selectlanguage{english}

\section{\label{sec:intro}Introduction}
Non-equilibrium continuous phase transitions have been and still are an object
of intense research activity. Underlying dynamic laws are responsible
for a more diverse behavior in contrast to the equilibrium models.
One of the most prominent examples is the directed bond 
 percolation \cite{HHL08,JanTau04} process that can be interpreted as
 a simple spreading process of an infection disease \cite{Janssen81}.  
 A distinguished feature is the presence of non-equilibrium phase transition
 between the active and absorbing state. There the correlation diverges and large scale
 properties become independent on the microscopic details. 
 It has been known that phase transitions are quite sensitive 
to additional disturbances such as quenched disorder \cite{Janssen96} or long-range
interactions \cite{Hinrichsen}. From practical point of view
this might be a reason why there are not so many experimental
realizations for the percolation process  \cite{TKCS07}.

In this paper we assume that the effect of environment can be simulated by advective velocity 
fluctuations with finite correlation in time and compressibility taken into account.
Following the works \cite{Ant99,Ant00} we  
 consider a generalization of the original Kraichnan model where the
advection-diffusion problem of non-interacting admixture has been studied. 
As some progress in this direction has already been made \cite{AntKap08,AntKap10,Ant11,DP13,DP15,DP16}, our main
aim here is to elucidate in detail the case of frozen velocity limit \cite{Ant00,HonKar88}.
  The main theoretical tool is the field-theoretic approach \cite{Vasiliev} with subsequent Feynman 
  diagrammatic technique and renormalization group (RG) approach, which allows us to determine the large-scale
 behavior, which is of special interest from macroscopic point of view.
 
 The paper is organized as follows. In 
 Sec.~\ref{sec:model}, we introduce a
field-theoretic version of the problem and in Sec.~\ref{sec:stable} we present an 
 analysis of possible regimes, which corresponds to the fixed points (FP) of the RG equations.
{\section{Field-theoretic model} \label{sec:model}}
The large scale properties of the directed bond percolation process \cite{HHL08,JanTau04} 
are captured by the following Langevin equation 
\begin{equation}
  \partial_t {\psi}  = D_0 (\boldnabla^2 - \tau_0)\psi  -   
   {\lambda_0 D_0}\psi^2/2 + \eta,
  \label{eq:langevin}
\end{equation}
where the field $\psi(t,\mx)$ corresponds to the density of percolating agents,
$\partial_t = \partial / \partial t$ is
the time derivative, $\boldnabla^2$ is  the Laplace operator, $D_0$ 
is the diffusion constant, $\lambda_0$ is the coupling constant and $\tau_0$ controls
 a deviation from the criticality.  The noise field $\eta(t,\mx)$ stems from the coarse
 graining procedure and mimics the influence of fast degrees of freedom. It is  assumed that
 $\eta$ is a random Gaussian variable with zero mean and correlator
\begin{equation}
  \langle \eta(t_1,\mx_1) \eta(t_2,\mx_2) \rangle = g_0 D_0 \psi(t_1,\mx_1) 
   \delta(t_1-t_2) \delta^{(d)}(\mx_1 - \mx_2),
  \label{eq:noise}
\end{equation}
where $\delta^{(d)}(\mx)$ is the $d$-dimensional delta function.
 The amplitude factor $\propto \psi$ 
 in (\ref{eq:noise}) ensures that the
fluctuations cease in the absorbing state $\psi= 0$.
 
 An advection process can be introduced in the standard fashion using
 the replacement of time derivative $ \partial_t$ by the Lagrangian derivative
 $\nabla_t \equiv \partial_t +(\mv\cdot\boldnabla)$. 
 A basic idea of the Kraichnan model \cite{Ant99,Ant00} is that
 the velocity field $\mv$ is a random Gaussian variable with zero mean and 
 a translationally invariant correlator \cite{Ant00} conveniently given
 in the Fourier representation as follows
\begin{equation}
  \langle v_i v_j \rangle_0 (\omega,\mk)
  = [P_{ij}^{k} + \alpha Q_{ij}^{k}] D_v(\omega,\mk),\quad
  D_v(\omega,\mk)=
  \frac{g_{10} u_{10} D_0^3 k^{4-d-y-\eta}}{\omega^2 + u_{10}^2 D_0^2 (k^{2-\eta})^2}.
  \label{eq:kernelD}
\end{equation}
Here, $P_{ij}^k = \delta_{ij}-k_ik_j/k^2$ is a transverse  and $Q_{ij}^k=k_ik_j/k^2$ a longitudinal
projection
operator. Further, the function $D_v$ is known as a kernel, where $k=|\mk|$ and a positive parameter $\alpha>0$ can be interpreted as a
deviation \cite{Ant00} from the incompressibility condition $\boldnabla\cdot\mv = 0$.
The coupling constant $g_{10}$ and the exponent $y$ describe the energy spectrum \cite{Ant99,Ant00} of the velocity
fluctuations. The constant $u_{10}>0$ and the exponent $\eta$ are related
to the characteristic frequency of the mode with wavelength $k$.
The kernel function for the correlator (\ref{eq:kernelD}) contains special cases: rapid-change model, frozen
 velocity ensemble and turbulence advection (see \cite{Ant99,Ant00}).
 Our aim here is to analyze the case
 of the frozen velocity limit, which is obtained in the limit $u_{10} \rightarrow 0$. It 
  bears resemblance to
a model of random walks in a random environment with long-range correlations \cite{HonKar88}.
The correlator (\ref{eq:kernelD}) attains then the form
	\begin{equation}
	   \langle v_i v_j \rangle_0 (\omega,\mk) \rightarrow
	   [P_{ij}^{k} + \alpha Q_{ij}^{k}] g_0 D_0^2 \pi \delta(\omega) k^{2-d-y}.
	   \label{eq:fvf_limit}
	\end{equation}
 In this limit velocity correlator does not depend on the time difference $(t-t')$.
The name ``frozen'' is motivated by the detailed analysis \cite{Ant99}, where it has been shown 
that in this limit the correlation time of the velocity field $t_v(k)$ as
a function of momentum scale $k$ is much larger than the correlation time of the advecting scalar quantity $t_\psi(k)$ and
effectively $t_v$ can be set to infinity. Hence from the point of view of $\psi$ field the velocity fluctuations
are not evolving in time, hence ``frozen''.

\begin{figure}[h!]
   \centering
   \begin{tabular}{c c}
     \includegraphics[width=5.6cm]{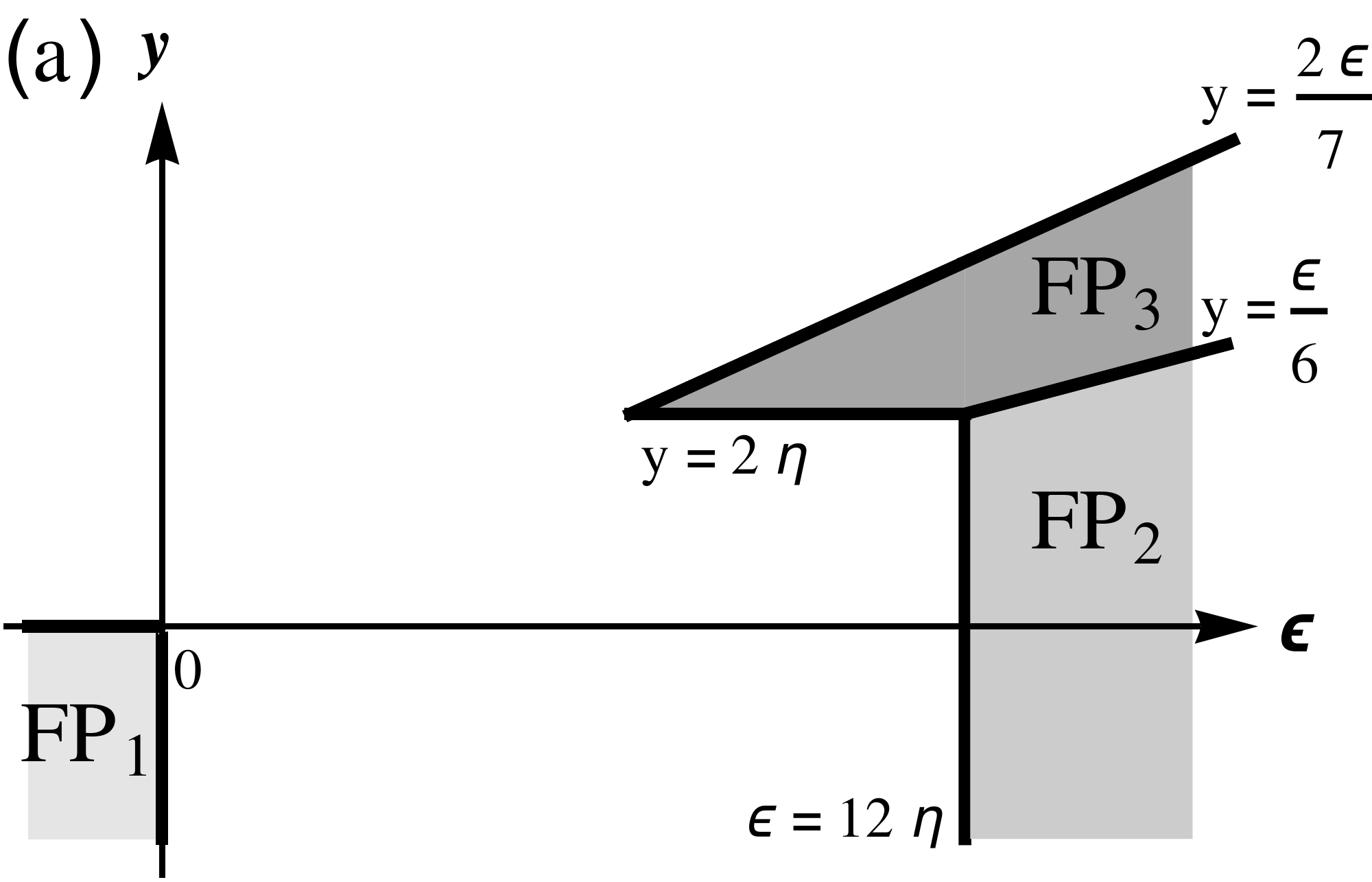}
     &
     \includegraphics[width=5.6cm]{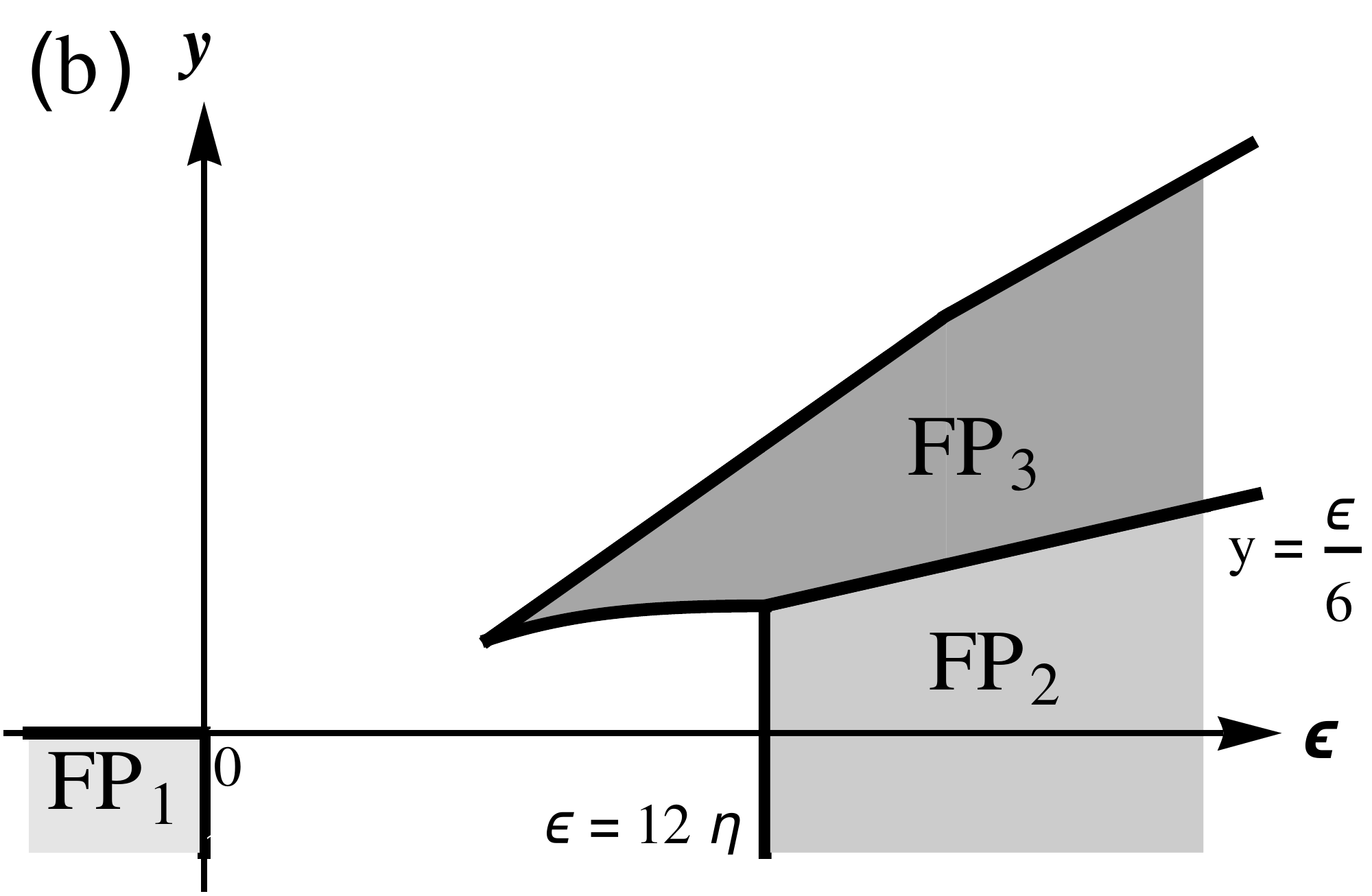}  \\
     \includegraphics[width=5.6cm]{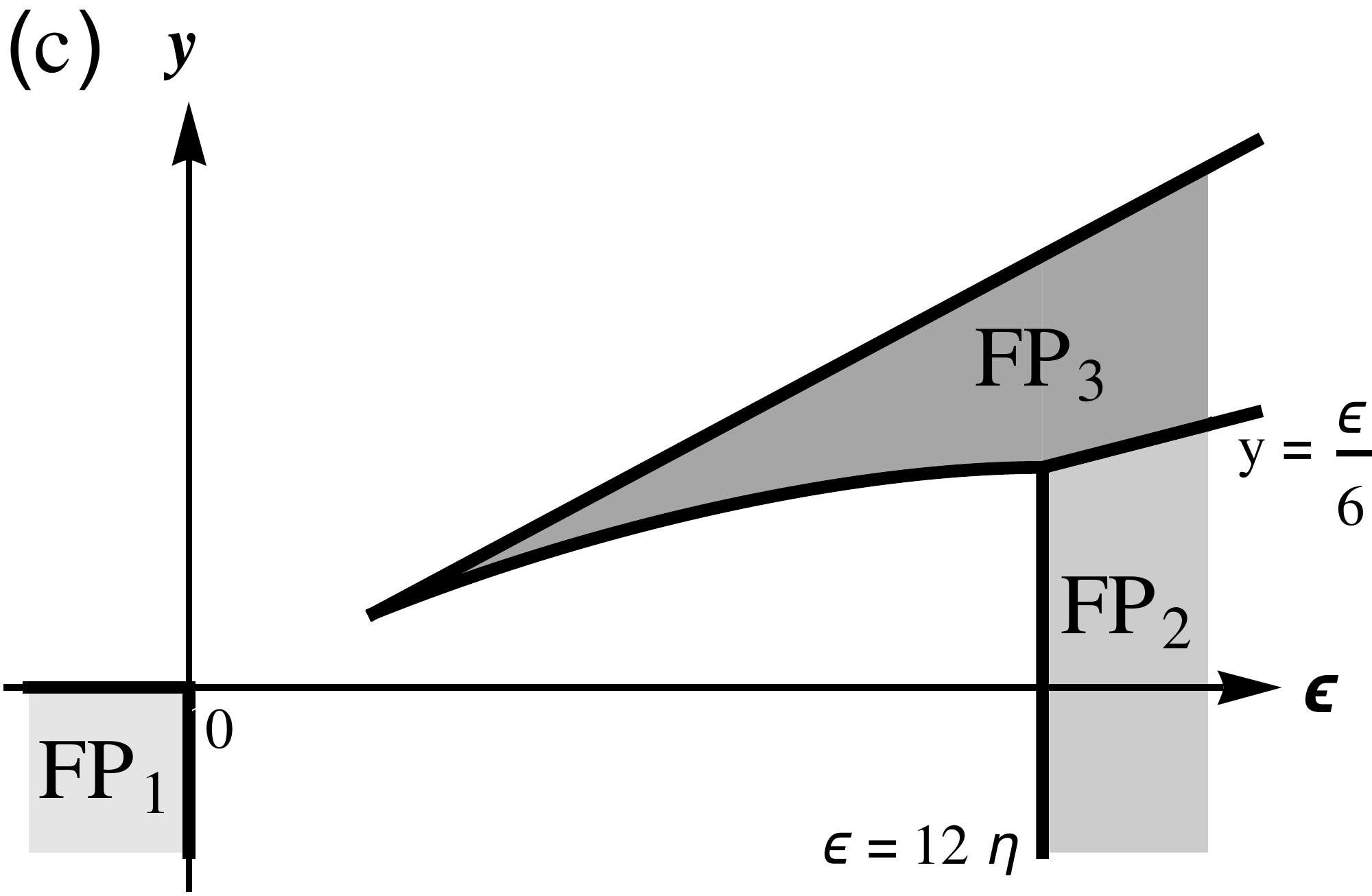}
     &
     \includegraphics[width=5.6cm]{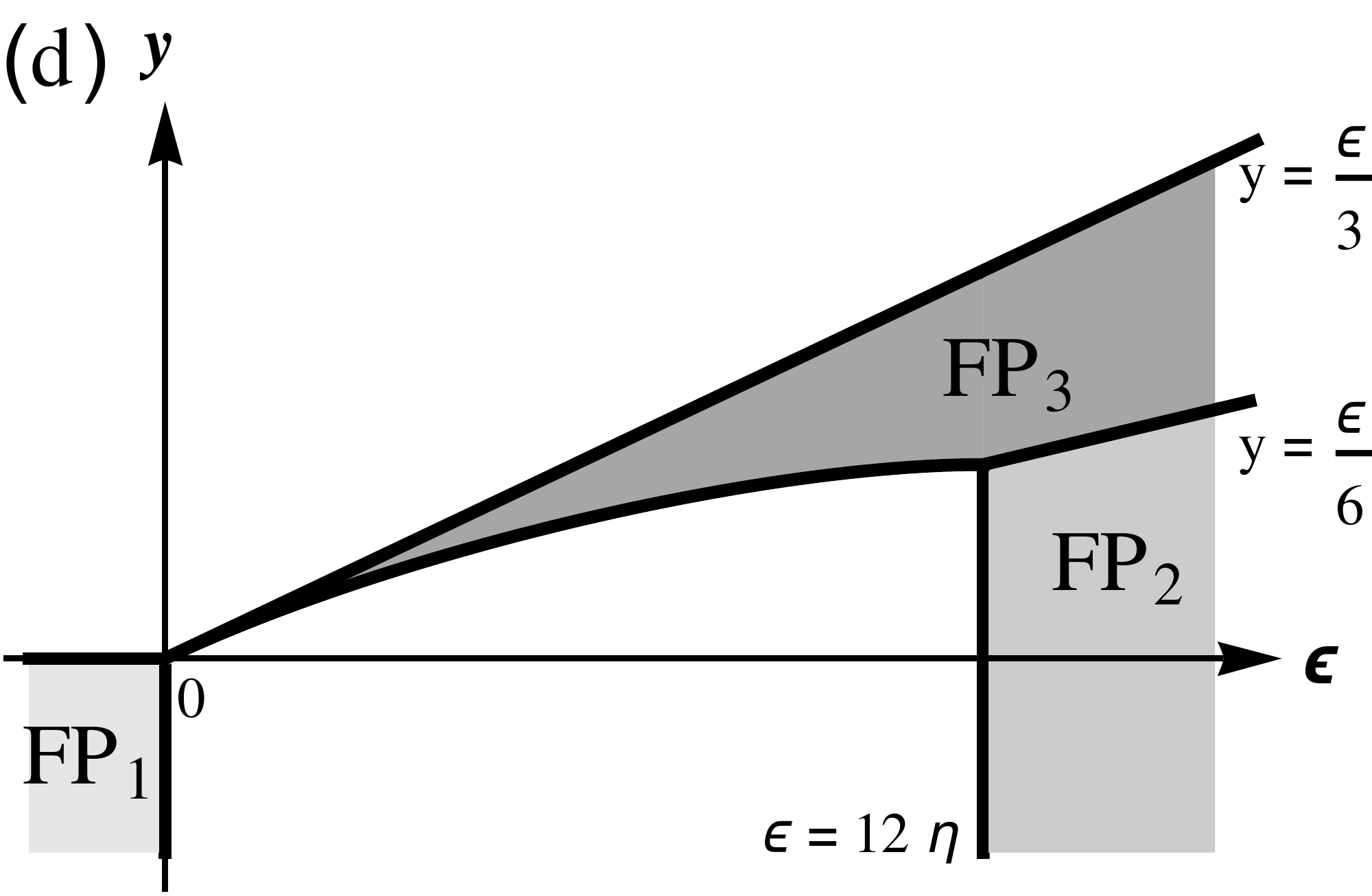}
      \end{tabular}
   \caption{Regions of stability in the plane $(\eps,y)$ for different choices of the compressibility parameter
   $\alpha$, which has been chosen to attain values $0$(a), $1$(b), $5$(c) and $6$(d).}
   \label{fig:frozen}
\end{figure}

The stochastic problem (\ref{eq:langevin}) and (\ref{eq:kernelD}) is amenable to the full machinery
of the quantum field theory methods as Feynman diagrammatic expansion and renormalization group method.
The latter can be done in a straightforward fashion following the well-established theory \cite{Vasiliev}
 and constitutes our main theoretical tool.
According to a general theorem \cite{Vasiliev}, the stochastic problem is equivalent to the field theoretic
model of a doubled set of fields $\tilde{\psi},\psi$ with Janssen-De Dominicis functional written
in a compact form as
\begin{align}
  \J[\tilde{\psi},\psi,\mv] & =  
  \tilde{\psi}[
  \nabla_t + D_0(\tau_0 -\nabla^2)
  ]\psi  + 
   {D_0\lambda_0} [\psi-\tilde{\psi}]\tilde{\psi}\psi/2 
   +  v D_v^{-1} v/2, \nonumber\\
   & + a_0 \tilde{\psi} ({\bm \nabla}\cdot{\bm v}) \psi - \frac{u_{20}}{2D_0} \tilde{\psi} \psi \mv^2.
  \label{eq:act_per}
\end{align}
We use a condensed notation in which integrals over the spatial variable 
$x$ and the time variable $t$ and summation over repeated index are implicitly assumed. 
For example, the first term on the right hand side actually stands for
$ \int \dRM t \int \dRM^{d} \mx\mbox{ } \tilde{\psi} (t,\mx)\allowbreak \partial_t \psi(t, \mx)$.
The last two terms in the action (\ref{eq:act_per}) have been added in order to ensure renormalizability
of the model. Their presence can be tracked down to the broken Galilean invariance
of the velocity ensemble and assumed compressibility. These directly lead to the more involved
numerical analysis of the flow equations than previously considered models \cite{AntKap08,AntKap10,Ant11,DP13}.
It can be proven \cite{DP16} that the field theoretic action (\ref{eq:act_per}) is closed with respect to
the renormalization group analysis and no other interactions than that already involved
in (\ref{eq:act_per}) are generated. Thus the model (\ref{eq:act_per}) is multiplicatively
renormalizable. Then according to the general theory of the RG group
one can exploit the correlation between ultraviolet and infrared (IR) singularities for a logarithmic theory and
 information about large-scale (macroscopic) behavior can be extracted.
In contrast to the standard $\eps$-expansion in
$\varphi^4$-theory we deal here with the three parameter expansion $(\eps,y,\eta)$, which
 are considered to be of the same order of magnitude. Here as usual $\eps=4-d$ is a deviation from
 the upper critical dimension $d_c=4$ \cite{JanTau04}.
{\section{Results of the RG analysis}\label{sec:stable}}
In the frozen velocity limit \cite{Ant00} the fixed point coordinate of the charge $u_1^*$ is zero.
The RG analysis reveals \cite{DP16} existence of eight possible regimes, from
which only three points \fp{1}, \fp{2} and \fp{3} are IR stable and hence possibly observable on
  the macroscopic scale.
 The physical
interpretation of the stable regimes is the following:
the fixed point \fp{1} describes the free (Gaussian) theory for which all interactions
are irrelevant and no scaling and universality is expected. The point \fp{2}
is a standard nontrivial percolation regime \cite{Janssen81} with irrelevant advection. The last point
\fp{3} embodies a nontrivial regime for which both velocity and
percolation interactions are relevant and corresponds to a new universality class. Because for points \fp{1} and \fp{2}
the velocity field is irrelevant, their boundaries are not affected by the value of the parameter
$\alpha$. On the other hand the stability region of \fp{3} heavily depends on it. 
For incompressible case $\alpha=0$ the boundaries between the regions can be computed exactly
and can be read out from Fig.~\ref{fig:frozen}(a).
From (\ref{eq:kernelD}) one can expect that the parameter $\eta$ formally drops out of the theory. However,
 RG analysis \cite{DP16} shows that $\eta$ affects the boundaries of fixed points.
  Note that in the blank space other regimes, e.g. rapid-change model
or turbulent diffusion, can be realized.
 
With increasing value of parameter
$\alpha$ situation becomes more involved. Numerical analysis shows that up to $\alpha=6$ 
the region of \fp{3} spreads to the origin, which can be seen on Figs.~\ref{fig:frozen} for 
different values of $\alpha$. Wherever it was possible, we have given the explicit form of boundary conditions.
The main observation is that compressibility in cooperation with percolation interactions can lead to a stabilization
of the nontrivial regime. 

\begin{figure}[h!]
   \centering
       \begin{tabular}{c c}
     \includegraphics[width=5.6cm]{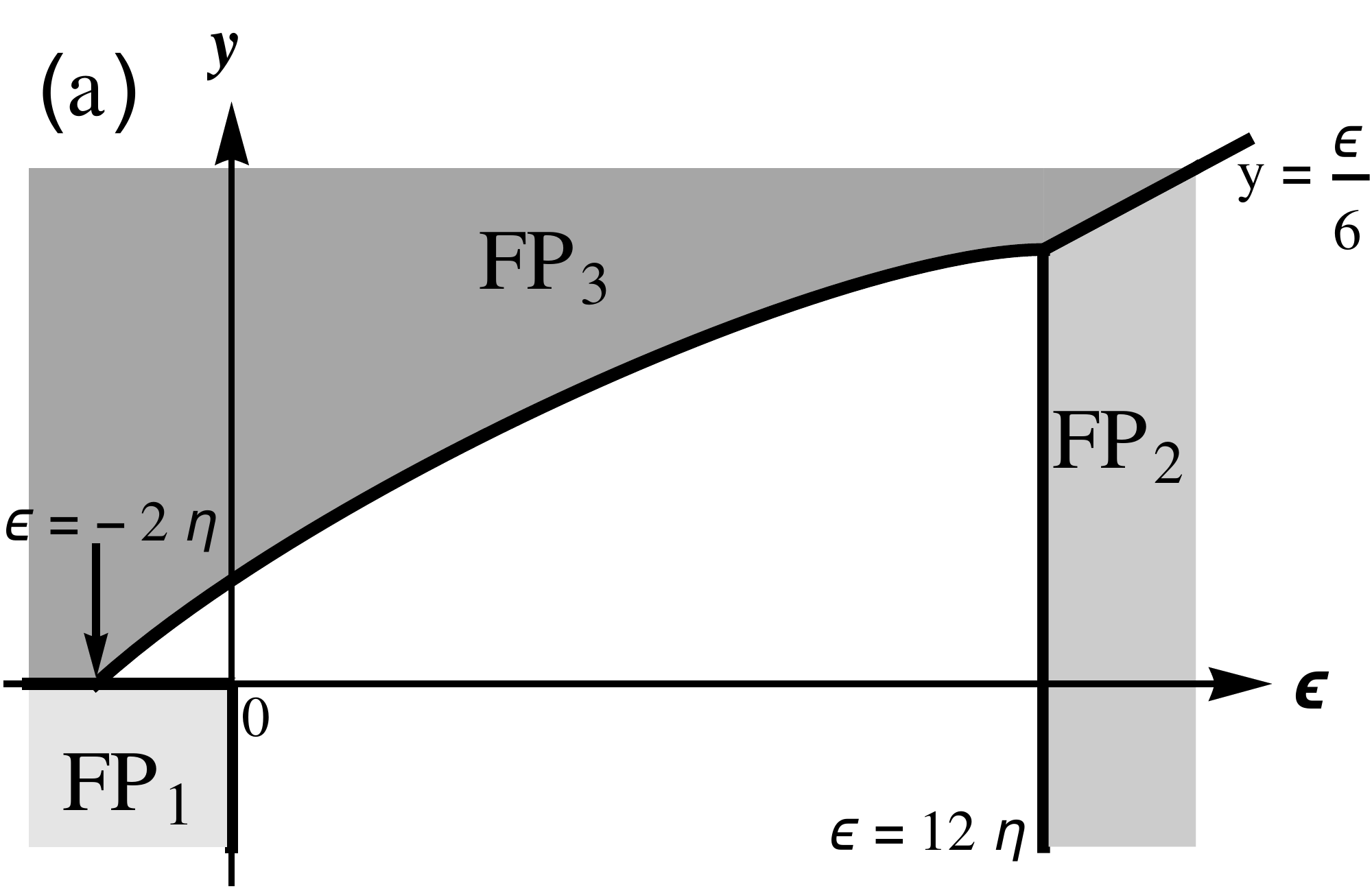}        
     &
     \includegraphics[width=5.6cm]{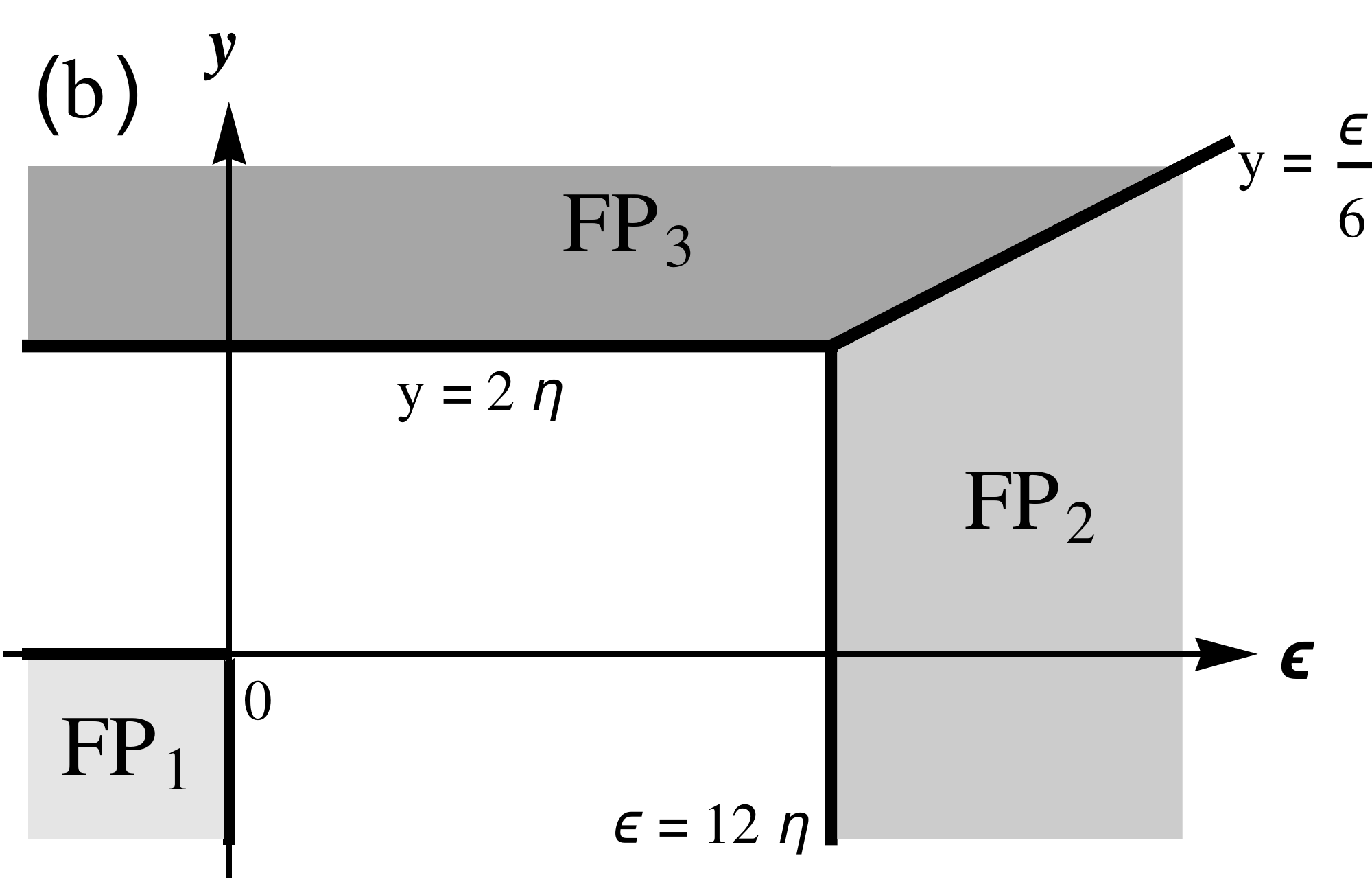}   
 \end{tabular}     
   \caption{Phase diagram obtained for $\alpha=12$ (a) and in the limit of pure irrotational velocity field 
   $\alpha\rightarrow\infty$ (b).}
   \label{fig:frozen2}
\end{figure}

For $\alpha\ge 6$ the lower boundary of the regime \fp{3} can be computed exactly and
is given by the following expression
\begin{equation}
   y = \frac{6 (\alpha-3) \epsilon + [\alpha (2 \alpha-21) - 54] \eta - 3 \sqrt{D} } {(\alpha-3) (\alpha+30)},
  \label{eq:boundary}
\end{equation}
where $D$ is the quadratic polynomial in the expansion parameters $\eta,\epsilon$
\begin{equation}
  D = [1764 + \alpha (49 \alpha - 372)] \eta^2 - 4 (\alpha - 42) (\alpha -3) \epsilon \eta  -12 (\alpha-3) \epsilon^2.
  \label{eq:boundary2}
\end{equation}
This situation is depicted on Fig.~\ref{fig:frozen2}. 
 With increasing value of $\alpha$  
the regime \fp{3} is stable also in the region with negative values of $\eps$ and the bottom
boundary is given by Eq. (\ref{eq:boundary}).
As \fp{3} can be IR stable also for $\eps<0$, which corresponds to the space dimensions $d>d_c=4$,
  we conclude that velocity fluctuations can destroy the expected mean-field behavior. 
  In the case of the pure irrotational (curl-free) field $\alpha\rightarrow\infty$ \ref{fig:frozen2}(b)
  the bottom boundary becomes a line $y=2\eta$.

From the theoretical point of view a special interest is paid to
the three-dimensional Kolmogorov regime \cite{Frisch}, which
is obtained for the choice $y=2\eta=8/3$ and $\eps=1$.
From Fig. \ref{fig:frozen2} it follows this regime
lies precisely on the boundary of the regime \fp{3}. According to detailed numerical analysis \cite{DP16} 
this regime competes with another nontrivial regime and exhibits non-universal behavior, i.e. dependence
on the initial value of bare parameters.
 {\section{Conclusions} \label{sec:conclu}}
In this paper we have studied percolation spreading in the presence of
 frozen velocity field with long-range correlations. The field theoretic formulation of the model has
 been formulated and the consequences following from its multiplicative
renormalizability have been discussed. 
We have found that depending on the values of a spatial dimension $d=4-\eps$, scaling
exponents $y$ and $\eta$, which describe statistics of velocity fluctuations, the
 model exhibits various universality classes.
Some of them are already well-known: the Gaussian (free) fixed point,
 a directed percolation without advection and  the last 
 one which corresponds to new universality classes, for which mutual interplay
between advection and percolation is relevant.
As has been shown \cite{GV99} the main drawback of the model is
that anomalous scaling behavior is destroyed when $\alpha$ and $y$ are large enough. Therefore
only relatively small values of $\alpha$ are admissible.
In order to investigate the case of strong compressibility and clarify the role of compressibility one should try to
 apply more sophisticated model for the velocity fluctuations, e.g. one considered in
  \cite{ANU97}. 
  
The work was supported by VEGA grant No. $1/0222/13$ 
 of the Ministry of Education, Science, Research and Sport of the Slovak Republic. N.~V.~A. and
  A.~S.~K. acknowledge Saint Petersburg State University for Research Grant No. 11.38.185.2014.
  A.~S.~K. was also supported by the grant 16-32-00086 provided by the Russian
Foundation for Basic Research.

\end{document}